\documentclass{jfm}

\usepackage{array}
\usepackage{graphicx}
\usepackage{subfigure}
\usepackage{amssymb}
\usepackage{amsmath} 
\usepackage{bm}
\usepackage{hyperref}
\usepackage{setspace}

\usepackage{lineno}

\newcommand*\patchAmsMathEnvironmentForLineno[1]{%
  \expandafter\let\csname old#1\expandafter\endcsname\csname #1\endcsname
  \expandafter\let\csname oldend#1\expandafter\endcsname\csname end#1\endcsname
  \renewenvironment{#1}%
     {\linenomath\csname old#1\endcsname}%
     {\csname oldend#1\endcsname\endlinenomath}}%
\newcommand*\patchBothAmsMathEnvironmentsForLineno[1]{%
  \patchAmsMathEnvironmentForLineno{#1}%
  \patchAmsMathEnvironmentForLineno{#1*}}%
\AtBeginDocument{%
\patchBothAmsMathEnvironmentsForLineno{equation}%
\patchBothAmsMathEnvironmentsForLineno{align}%
\patchBothAmsMathEnvironmentsForLineno{flalign}%
\patchBothAmsMathEnvironmentsForLineno{alignat}%
\patchBothAmsMathEnvironmentsForLineno{gather}%
\patchBothAmsMathEnvironmentsForLineno{multline}%
}

\shorttitle{Scaling of the Puffing Strouhal Number for Buoyant Jets}
\shortauthor{N. T. Wimer et al.}

\title{Scaling of the Puffing Strouhal Number for Buoyant Jets}

\author{N.~T.~Wimer\aff{1}\corresp{\email{Nicholas.Wimer@Colorado.edu}}, C.~Lapointe\aff{1}, J.~D.~Christopher\aff{1}, S.~P.~Nigam\aff{1}, T.~R.~S.~Hayden\aff{1}, A.~Upadhye\aff{2}, M.~Strobel\aff{2}, G.~B.~Rieker\aff{1} \and P.~E.~Hamlington\aff{1}}

\affiliation{
\aff{1}Department of Mechanical Engineering, University of Colorado, Boulder, CO 80309, USA
\aff{2}Corporate Research Process Laboratory, 3M Company, St Paul, MN 55144, USA
}

\begin{document}

\maketitle

\begin{abstract}
Prior research has shown that round and planar buoyant jets ``puff'' at a frequency that depends on the balance of momentum and buoyancy fluxes at the inlet, as parametrized by the Richardson number. Experiments have revealed the existence of scaling relations between the Strouhal number of the puffing and the inlet Richardson number, but geometry-specific relations are required when the characteristic length is taken to be the diameter (for round inlets) or width (for planar inlets). In the present study, we show that when the hydraulic radius of the inlet is instead used as the characteristic length, a single Strouhal-Richardson scaling relation is obtained for a variety of inlet geometries. In particular, we use adaptive mesh numerical simulations to measure puffing Strouhal numbers for circular, rectangular (with three different aspect ratios), triangular, and annular high-temperature buoyant jets over a range of Richardson numbers. We then combine these results with prior experimental data for round and planar buoyant jets to propose a new scaling relation that accurately describes puffing Strouhal numbers for various inlet shapes and for Richardson numbers spanning over four orders of magnitude.
\end{abstract}

\begin{keywords}
\end{keywords}

\section{Introduction\label{sec:intro}}

Buoyant jets are found in a variety of natural and engineering contexts, including industrial burners \citep{Christopher2018,Hayden2019}, hydrothermal vents \citep{Gaskin2001}, and volcanic plumes \citep{Campion2018}. Despite their obvious differences, each of these flows exhibits a similar ``puffing'' phenomenon that results from the continuous injection of less dense fluid into a reservoir of more dense, ambient fluid. The associated flux of both momentum and buoyancy through the inlet leads to the formation of vortical structures that are periodically shed, yielding the puffing behavior. The resulting puffing frequency, $f$, can be characterized using the Strouhal number, $\mathrm{St}_\ell=f\ell/V_0$, and the balance between momentum and buoyancy inlet fluxes is characterized by the Richardson number, $\mathrm{Ri}_\ell =(1-\rho_0/\rho_\infty)g\ell/V_0^2$, where $\ell$ and $V_0$ are, respectively, the characteristic dimension and inlet velocity of the jet, $\rho_0$ is the density of the inlet fluid, $\rho_\infty$ is the ambient density, and $g$ is the gravitational acceleration.

Buoyant jet puffing has been examined in detail over the past three decades, beginning with \citet{Hamins1992} and \citet{Cetegen1996}. In both of these studies, round helium buoyant jets were studied over a range of Richardson and Reynolds, $\mathrm{Re}_\ell = V_0\ell/\nu_0$, numbers, where $\nu_0$ is the kinematic velocity of the inlet fluid. \citet{Hamins1992} found that the Strouhal and Richardson numbers were related as $\mathrm{St}_D \sim \mathrm{Ri}_D^{0.38}$, where $D$ is the inlet diameter. \citet{Cetegen1996} similarly determined the relation $\mathrm{St}_D = 0.8\mathrm{Ri}_D^{0.38}$ for $\mathrm{Ri}_D < 100$ and $\mathrm{St}_D = 2.1\mathrm{Ri}_D^{0.28}$ for $100 < \mathrm{Ri}_D < 500$. In a study of planar (i.e., high aspect ratio rectangular) buoyant jets, \citet{Cetegen1998} found the alternative relation $\mathrm{St}_W = 0.55\mathrm{Ri}_W^{0.45}$, where $W$ is the inlet width. The differences in round and planar buoyant jet scaling relations were attributed to differences in mixing rates and buoyancy fluxes between the different inlet shapes. Additionally, the puffing Strouhal number was found to be independent of Reynolds number \citep{Cetegen1997}.

Although these scaling relations are useful when studying round or planar buoyant jets, they are not directly applicable to inlets of different shapes. This can pose challenges in many practical applications involving complex inlets (e.g., industrial burners or volcanic plumes). Recently, \citet{Bharadwaj2019} experimentally studied a series of buoyant jets from rectangular inlets with different aspect ratios, and found that all puffing Strouhal numbers could be described using a single Richardson number scaling relation when the characteristic length, $\ell$, was taken to be the hydraulic diameter of the inlet, and the characteristic velocity was taken to be the effective velocity of a jet with the same total mass flow rate issuing through a round exit with a diameter equal to the hydraulic diameter. However, the question remains whether there exists a single universal scaling relation based on the hydraulic diameter (or radius) that can be used to relate Strouhal and Richardson numbers for all inlet geometries, and not just rectangles with different aspect ratios.

In the present study, we take the first steps towards determining whether such a universal relation exists by using adaptive mesh numerical simulations to examine high-temperature buoyant jets for circular, rectangular (with three different aspect ratios), triangular, and annular buoyant jet inlet geometries. A range of Richardson numbers are examined for each geometry, and we show that a single scaling relation based on the hydraulic radius accurately describes all available experimental and computational data for Richardson numbers spanning over four orders of magnitude, even without the use of the effective velocity proposed by \citet{Bharadwaj2019}.  

\section{Description of Adaptive Mesh Numerical Simulations}
The numerical code used to perform the adaptive mesh simulations is \texttt{PeleLM}, a low-Mach reacting flow code \citep{Almgren1998,Day2000,Bell2005,Nonaka2012,Nonaka2018} that solves the Navier-Stokes equations, as well as additional equations for enthalpy and species conservation. All fluid transport properties are calculated as mixture-averaged coefficients assuming a mixture of perfect gases. A second-order Godunov scheme is used for advection and a semi-implicit discretization is used for diffusion. The overall numerical method is second-order accurate in both space and time, and the time-step is dynamically determined according to an advective Courant-Friedrichs-Levy condition. A more comprehensive description of the numerical approach, algorithm, and methods is available in \citet{Nonaka2018} and \citet{Wimer2019}. 
 
The simulations were each performed in a 1~m$^3$ domain using adaptive mesh refinement (AMR) to reduce the computational cost. The adaptive grid approach implemented in \texttt{PeleLM} is described in detail by \citet{Day2000}, and the governing equations were solved on a series of uniform, nested grids with no subgrid-scale modeling. The computational domain for each simulation consisted of a 128$^3$ base grid with two levels of AMR. The grid was re-meshed every time step to maintain sufficiently fine resolution in regions with large vorticity and density gradient magnitudes, and to reduce resolution in regions where magnitudes were small. With the two levels of AMR, the smallest physical scale resolved in the simulations was 1.95~mm, which was recently shown to be sufficiently fine for accurately capturing the puffing frequency of a large-scale helium plume using an identical numerical approach and similar physical setup \citep{Wimer2019}.

The bottom boundary in each of the simulations consisted of a high temperature jet inlet and a mild air co-flow. Open boundary conditions were used on the four sides and at the top of the domain. The jet inlet and co-flow were specified using Dirichlet boundary conditions, and the open boundaries were specified using a divergence-constrained velocity projection that allowed for both fluid inflow and outflow. In each of the simulations, hot air with temperature $T_0 = 1000$~K flowed through the jet inlet, and the inlet velocity, $V_0$, was varied to span a range of Richardson numbers (see Table \ref{tab:sims}). The co-flow was uniform and the same in all simulations, with velocity $V_\text{coflow} = 0.05$~m/s and temperature $T_\text{coflow} = 300$~K. The gravitational acceleration was directed downwards with magnitude $g=9.81$~m/s$^2$ towards the bottom boundary along the vertical axis.

\begin{table}
  \begin{center}
\begin{tabular}{cccccccc}
Inlet Shape         & Dimensions                    & $AR$  & Velocities $V_0$ (m/s)\\[3pt]
Circle              & 0.154~m (diameter)            & --    & 0.25                  \\
Circle              & 0.274~m (diameter)            & --    & (0.125, 0.25, 0.5, 1) \\
Rectangle           & $0.137\times 0.137$~m$^2$     & 1     & (0.25, 0.5, 1)        \\
Rectangle           & $0.075\times 0.25$~m$^2$      & 10/3  & (0.125, 0.25, 0.5, 1) \\
Rectangle           & $0.0433\times 0.433$~m$^2$    & 10    & (0.125, 0.25)         \\
Equil.\ Triangle    & $0.208$~m (side)              & --    & (0.125, 0.25, 0.5, 1) \\
Annulus             & 0.1755/0.234~m (in/out)       & 3/4   & (0.25, 0.5, 1, 1.5)   \\
\end{tabular}
\caption{Summary of simulations performed, indicating the shape of the inlet, its dimensions and aspect ratio ($AR$), and the inlet velocities, $V_0$, examined for each shape. In all simulations, the inlet temperature was $T_0 = 1000$~K and the 1~m$^3$ domain was discretized with a $128^3$ base grid and two levels of AMR, providing an effective grid resolution of 1.95~mm.}
\label{tab:sims}
  \end{center}
\end{table}

Seven different jet inlet shapes were simulated (as also summarized in Table \ref{tab:sims}): circles with two different diameters, rectangles with three different aspect ratios, an equilateral triangle, and an annulus. The aspect ratios of the rectangles are computed as  $AR_\text{rect}=L/W$, where $L$ is the length and $W$ is the width, with $L>W$ and $1 \le AR_\text{rect} < \infty$. Three rectangular aspect ratios were considered here, corresponding to $AR_\text{rect} = 1,10/3,10$. The aspect ratio for the annulus is given by $AR_\text{annu} = D_\text{in}/D_\text{out}$, where $D_\text{in}$ is the inner diameter and $D_\text{out}$ is the outer diameter, with $D_\text{in} < D_\text{out}$ and $0 \le AR_\text{annu} < 1$. An annulus of aspect ratio, $AR_\text{annu} = 3/4$ was examined here. The simulations were all performed using inlets of equal area and inlet temperature, ensuring that the momentum and buoyancy fluxes were identical for all simulations with the same inlet velocity, $V_0$. Each of the simulations were performed for 20~s and statistics were computed over the last 10~s to allow for the decay of initial transients. 

\section{Results}

\begin{figure}
\centering\includegraphics[width=\textwidth]{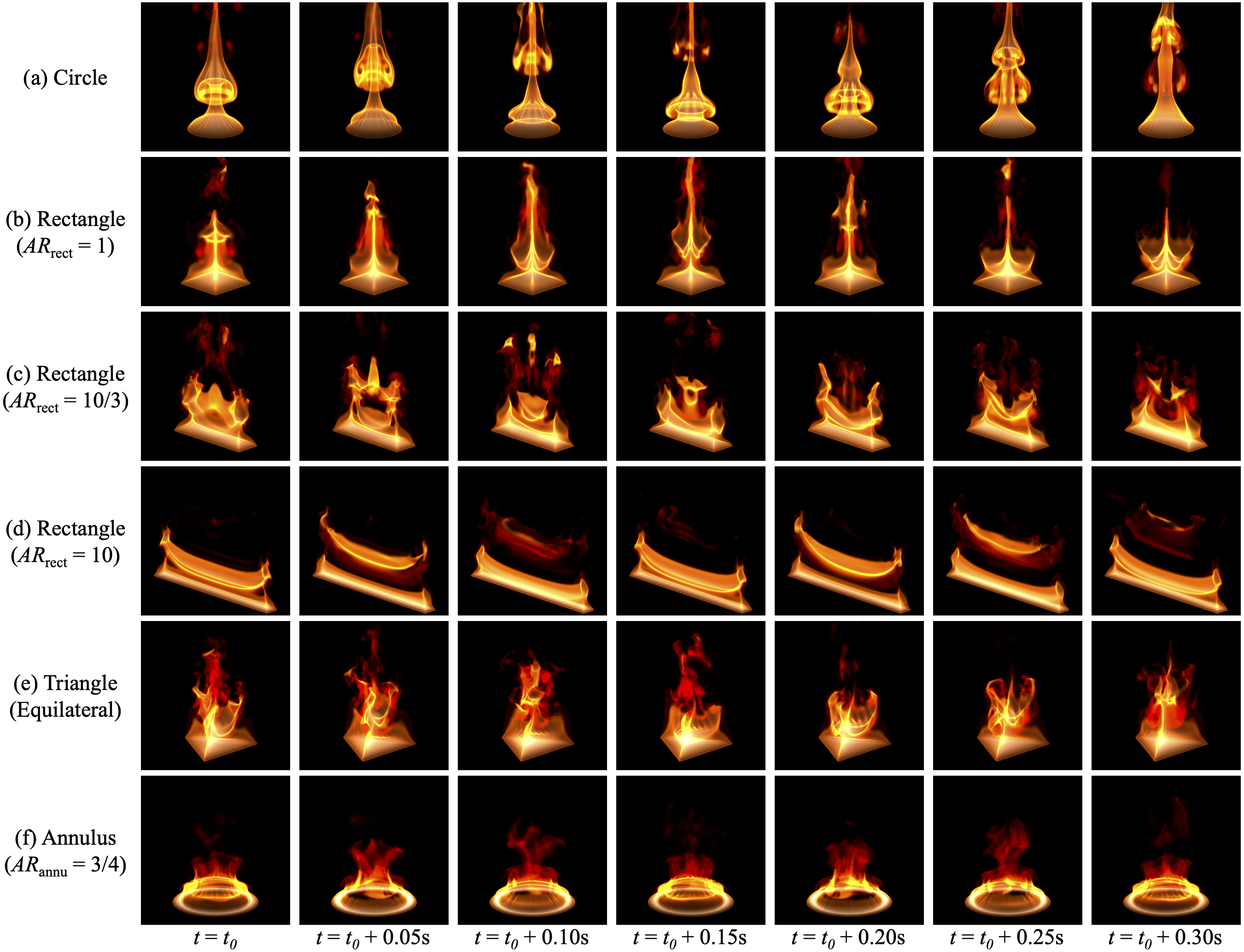}
\caption{Isosurfaces of the temperature field as a function of time (snapshots are separated by 0.05~s) for (a) a circular buoyant jet with $D = 0.154$~m, (b-d) rectangular buoyant jets with $AR_\mathrm{rect} = 1,10/3,10$, (e) an equilateral triangle buoyant jet with side $0.208$~m, and (f) an annular buoyant jet with $AR_\mathrm{annu} =3/4$. The inlet velocity in each case was $V_0=0.25$~m/s.}
\label{fig:volumes}
\end{figure}

Figure~\ref{fig:volumes} shows time series of temperature isosurfaces for each of the inlet geometries, with $V_0=0.25$~m/s in each case. For the circular inlet in Figure \ref{fig:volumes}(a), the hot air entering the domain is accelerated upwards along the centerline due to buoyancy. This subsequently causes entrainment of more dense cold air from the sides, pinching the hot gases towards the center of the inlet and creating a shear layer between the hot gases and cooler ambient air. This shear layer, combined with the difference in density across the layer, leads to an axisymmetric Kelvin-Helmholtz (KH) instability around the circumference of the inlet. This instability rolls up into a toroidal vortex ring that eventually pinches off and is shed, as indicated by the sequence of toroidal temperature isosurfaces rising above the inlet in Figure \ref{fig:volumes}(a).

Figure~\ref{fig:volumes}(b) shows that the flow evolution for an $AR_\text{rect}=1$ rectangular inlet is similar to that of the circular inlet, although the resulting structures differ. Hot air is still accelerated upwards upon entering the domain due to the buoyant force, again leading to entrainment of cold air from the sides and pinching of the hot gases towards the centroid of the square. However, the entrained flow along the sides of the inlet reaches the centroid before the flow originating from the corners. This leads to the roll-up and shedding of vortical structures (again indicated by the temperature isosurfaces) that are less coherent than in the circular case shown in Figure \ref{fig:volumes}(a). 

As the aspect ratio of the rectangle increases to $AR_\text{rect}=10/3$ and $10$, shown in Figures~\ref{fig:volumes}(c) and (d), respectively, the shed vortical structures become increasingly elongated. Since the aspect ratio is greater than one in these cases, the long sides of the rectangle are closer to the centroid than the short sides and, as a result, the long-sided shear layer pinches off before the short-sided shear layer has time to develop. This causes the shedding of elongated vortices that are reminiscent of stretched vortex rings.

For the equilateral triangle inlet in Figure~\ref{fig:volumes}(e), the entrained flow from the sides reaches the centroid before the flow from the corners. This results in the formation of a three-sided vortical structure similar to the four-sided vortical structure of the square buoyant jet in Figure~\ref{fig:volumes}(b).

Finally, Figure~\ref{fig:volumes}(f) shows the structure and development of the buoyant jet for the annular inlet. As the hot air enters the domain, ambient air is again entrained, not towards the centerline of the shape, but rather towards the average of the inner and outer radii (i.e., directly above each section of the inlet). This location of converging flow and subsequent puffing causes ambient air to be entrained in the negative radial direction at locations greater than the outer radius, and in the positive radial direction at locations less than the inner radius. Consequently, the shear layer, roll-up, and subsequent shedding of vortical structures are all centered above the average of the inner and outer radii. This results in toroidal vortices within the inner radius that create a net downward motion at the center of the annulus and counteract the upward acceleration of fluid due to buoyancy.

For each of the inlet shapes examined here, the sequence of vertical buoyancy-driven flow, entrainment of ambient air, pinching of hot gases, development of the KH instability, and subsequent shedding of vortical structures is repeated indefinitely, resulting in characteristic puffing frequencies for each case. From a quantitative perspective, the dominant puffing frequency can be determined using a number of methods, and here we use a Fast Fourier Transform (FFT) of the vertical velocity time series at a point above the centroid of the circular, rectangular, and triangular buoyant jets, and above the average radius for the annular buoyant jet. The dominant frequency was found not to depend on height above the inlet, and a height equal to the hydraulic radius $R_\mathrm{h}=A/P$ was thus used for each case, where $A$ is the geometric area of the inlet and $P$ is its perimeter.

\begin{figure}
\centering\includegraphics[width=\textwidth]{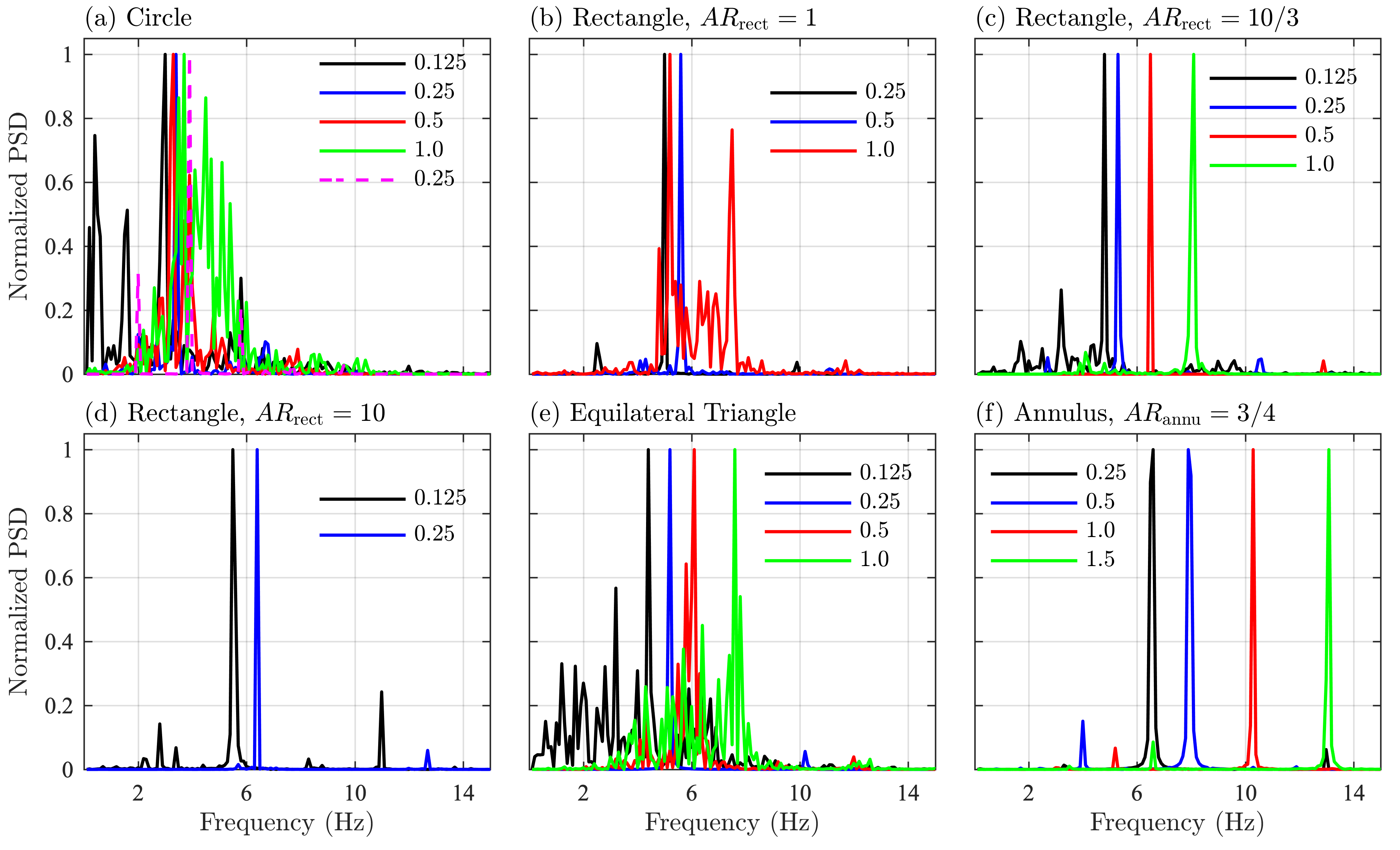}
\caption{Power spectral densities (PSDs) of the vertical velocity for (a) circular, (b-d) rectangular (with $AR_\text{rect} = 1,10/3,10$), (e) triangular, and (f) annular buoyant jets. Inlet velocities, $V_0$, are indicated in the legends in units of m/s. In panel (a), solid lines denote results for the $D=0.274$~m inlet and the dashed line corresponds to the $D=0.154$~m inlet. PSDs are computed one hydraulic radius above the inlet centroid for the circular, rectangular, and triangular inlets, and above the average radius for the annular inlet.}
\label{fig:ffts}
\end{figure}

The resulting power spectral densities, shown in Figure~\ref{fig:ffts}, display distinct peaks for each of the inlet shapes and velocities. Peak frequencies increase with inlet velocity for all shapes, and a comparison of the three rectangular cases in Figures~\ref{fig:ffts}(b-d) shows that the peak frequencies also increase as the aspect ratio increases. The highest peak frequencies are obtained for the annular case shown in Figure~\ref{fig:ffts}(f). Notably, despite the self-interacting nature of the annular buoyant jet and the subsequent difference in structure compared to the other inlet shapes, the flow still exhibits a characteristic puffing behavior. This indicates that the puffing phenomenon is largely independent of the inlet shape, and is instead determined by the underlying dynamics associated with buoyancy, entrainment, pinching, KH roll-up, and vortex shedding.

\section{Scaling and Universality}
Three non-dimensional groups can be formed from the dominant puffing frequency $f$ and the parameters that describe the buoyant jet inlet (i.e., $V_0$, $\ell$, $\rho_0$, and $\nu_0$); namely, $\mathrm{St}_\ell$, $\mathrm{Ri}_\ell$, and $\mathrm{Re}_\ell$. Prior work on round and planar buoyant jets has shown that $\mathrm{St}_\ell$ is independent of $\mathrm{Re}_\ell$ \citep{Cetegen1997} and, following previous studies, we thus seek a scaling relation of the form $\mathrm{St}_\ell=b\mathrm{Ri}_\ell^a$, where $a$ and $b$ are constants that will be determined from the present simulations and prior experimental data.

It is worth noting that, in order for $f$ to increase with inlet velocity $V_0$ (as is suggested for all inlet shapes by Figure \ref{fig:ffts}), the exponent $a$ must be smaller than $1/2$. Prior empirical scaling relations have all been consistent with this constraint; for round buoyant jets, \citet{Cetegen1996} found $a=0.38$ for $1 < \mathrm{Ri}_D < 100$ and $a=0.28$ for $\mathrm{Ri}_D > 100$, and for planar buoyant jets, \citet{Cetegen1998} found  $a=0.45$ for $1 < \mathrm{Ri} < 100$, with no secondary scaling relation reported.

Prior scaling relations between $\mathrm{St}_\ell$ and $\mathrm{Ri}_\ell$ have primarily used a characteristic length scale $\ell$ associated with the diameter or width of the inlet. As an initial step, here we follow a similar approach and examine the scaling of $\mathrm{St}_\ell$ using width-based length scales for each inlet shape. For the circular inlets, $\ell$ is taken to be the diameter, $D$. For the three rectangular inlets, the smallest side length (i.e., the width, $W$) is used for $\ell$, since this is the scale most relevant to the pinching and subsequent shedding of vortical structures. For equilateral triangles, similar to squares, the length of one side is used for $\ell$. Finally, for the annulus, $\ell$ is taken to be the difference between the outer and inner radii; i.e., $\ell=(D_\text{out} - D_\text{in})/2$.

\begin{figure}
\centering\includegraphics[width=\textwidth]{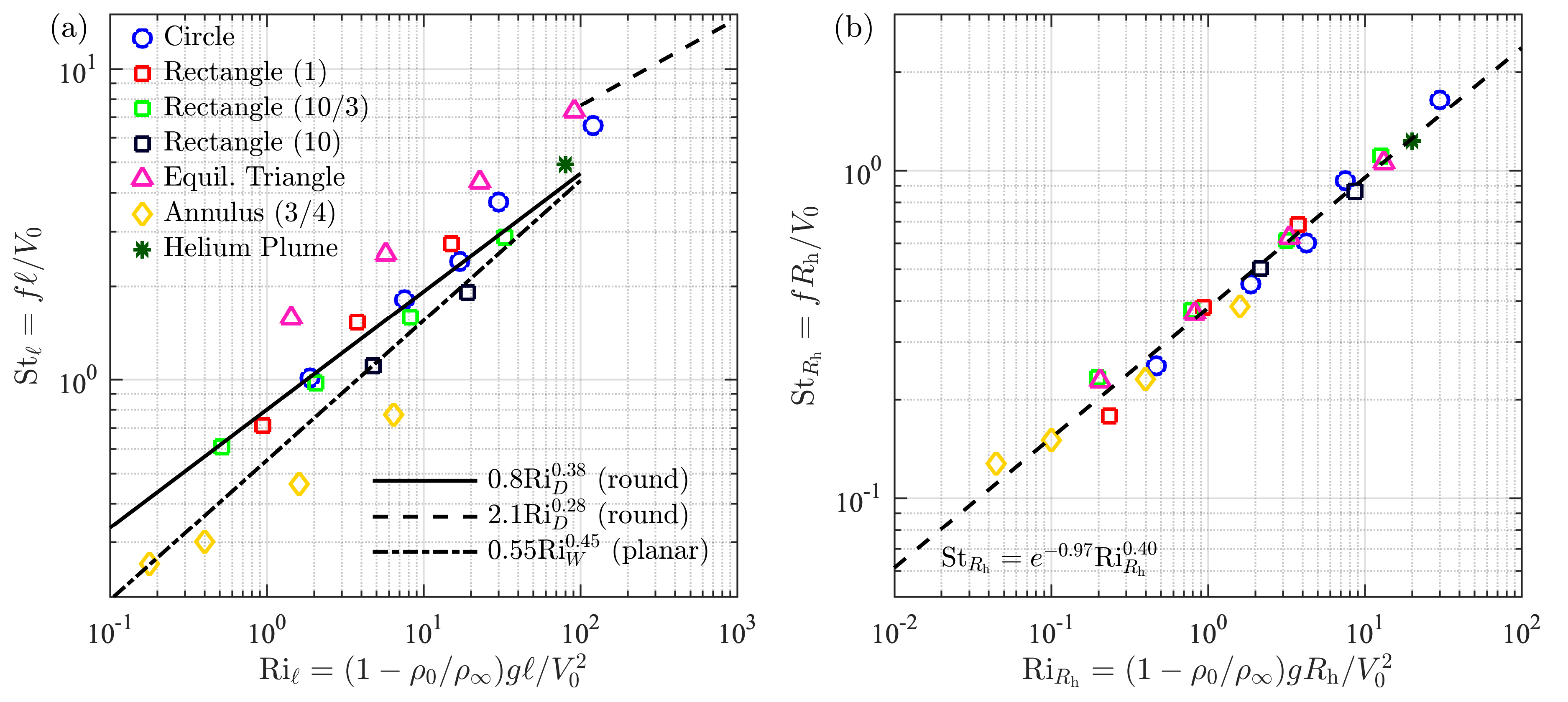}
\caption{Strouhal number as a function of Richardson number for circular, rectangular ($AR_\mathrm{rect} =1,10/3,10$), triangular (equilateral), and annular ($AR_\mathrm{annu} = 3/4$) buoyant jets. Panel (a) shows results for characteristic lengths $\ell$ based on the diameter (circles), width (rectangles), side (triangles), and difference between outer and inner radii (annuli). Panel (b) shows results when the characteristic length is taken to be the hydraulic radius, $R_\mathrm{h}$. Scaling relations from \citet{Cetegen1996} and \citet{Cetegen1998} are also shown in panel (a).}
\label{fig:StVRiTraditional}
\end{figure}

Figure~\ref{fig:StVRiTraditional}(a) shows $\mathrm{St}_\ell$ versus $\mathrm{Ri}_\ell$ for each of the simulations performed in the present study, as well as for the round helium plume case from \citet{Wimer2019}. Scaling relations from prior experimental studies of round \citep{Cetegen1996} and planar \citep{Cetegen1998}  buoyant jets are also shown. The circular buoyant jet results follow the round jet scaling relations from \citet{Cetegen1996}, including the transition at $\mathrm{Ri}_D \approx 100$. The rectangular buoyant jet results fall between prior round and planar scaling relations, with the lower aspect ratio rectangles more closely resembling the round scaling and the higher aspect ratio rectangles more closely resembling the planar scaling. The latter correspondence occurs because, as the rectangle becomes longer relative to its width (i.e., as aspect ratio increases), locations above the centroid appear more and more two-dimensional as corner effects become increasingly negligible. The triangular buoyant jet results follow a similar relation to that for the circular buoyant jets, but with larger values of $\mathrm{St}_\ell$. Interestingly, the annular buoyant jets also closely follow the round jet scaling slope, but with smaller values of $\mathrm{St}_\ell$.

Although scaling relations for each of the different inlet shapes could be obtained separately using least-squares fits to the data in Figure \ref{fig:StVRiTraditional}(a), we instead seek a single scaling relation between appropriately defined Strouhal and Richardson numbers for any inlet shape. In particular, here we introduce Strouhal and Richardson numbers, denoted $\mathrm{St}_{R_\mathrm{h}}$ and $\mathrm{Ri}_{R_\mathrm{h}}$, respectively, that use the hydraulic radius of the inlet, $R_\mathrm{h}$, as the characteristic length $\ell$. As an example, the hydraulic radius of a circle with diameter $D$ is $R_\mathrm{h} = \pi (D/2)^2 /(\pi D) = D/4$, and the functional forms of $R_\mathrm{h}$ for other inlet shapes are similarly straightforward to determine.

It should be noted that there is substantial reason to suspect the importance of the hydraulic radius in the formulation of a geometry-independent scaling relation for the puffing Strouhal number. In particular, \citet{Bharadwaj2019} recently showed that Strouhal and Richardson numbers based on the hydraulic diameter resulted in a single scaling relation for a series of rectangular buoyant jets with different aspect ratios. From a physical standpoint, the inlet area and perimeter used to determine the hydraulic radius are also dynamically significant; the area determines, in part, the total buoyancy flux introduced into the domain, and the shear layer that leads to KH roll-up and subsequent vortex shedding forms along the inlet perimeter.   

Figure~\ref{fig:StVRiTraditional}(b) shows the resulting relationship between $\mathrm{St}_{R_\mathrm{h}}$ and $\mathrm{Ri}_{R_\mathrm{h}}$ for each of the inlet shapes and velocities. Most notably, all results now fall close to a single scaling relation that was determined using a least-squares fit to be $\mathrm{St}_{R_\mathrm{h}}=e^{-0.97}\mathrm{Ri}_{R_\mathrm{h}}^{0.40}$, with an $r^2 = 0.981$ coefficient of correlation. The secondary scaling associated with circular buoyant jets is eliminated, and all round buoyant jets, including the helium plume simulation from \citet{Wimer2019}, follow the same scaling relation. Additionally, the various rectangular, triangular, and annular results all also follow this relation.

\begin{figure}
\includegraphics[width=\textwidth]{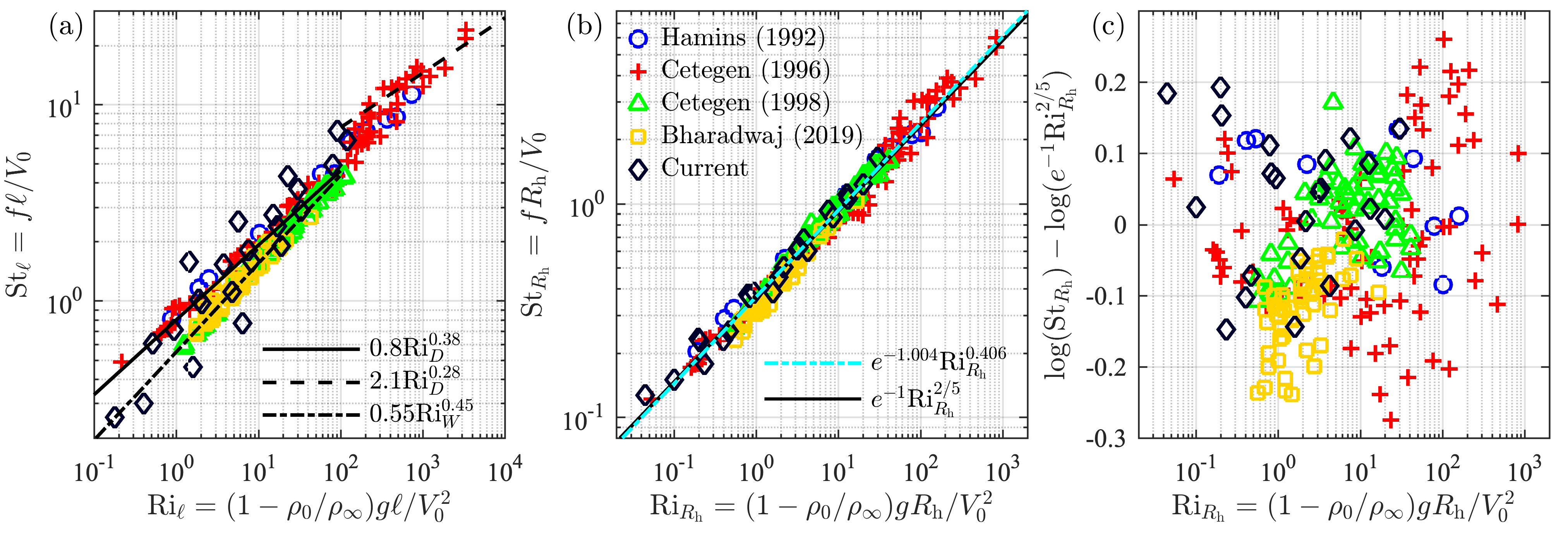}
\caption{Strouhal number as a function of Richardson number for the present simulations and prior experiments from \citet{Hamins1992}, \citet{Cetegen1996}, \citet{Cetegen1998}, and \citet{Bharadwaj2019} [see legend in panel (b)]. Definitions of $\ell$ in (a) and (b) are the same as those in Figure~\ref{fig:StVRiTraditional}. In panel (b), we show the least squares fit to all data, as well as the proposed relationship from Eq.\ \eqref{eq:law}. In panel (c), we show residuals with respect to Eq.\ \eqref{eq:law}.}
\label{fig:StVRiHydraulicRadius}
\end{figure}

The single scaling relation measured using the present simulations can also be extended to include prior experimental data for round, planar, and rectangular inlets. In particular, Figure~\ref{fig:StVRiHydraulicRadius} shows data from the present simulations along with all prior experimental data from \citet{Hamins1992}, \citet{Cetegen1996}, and \citet{Cetegen1998}, as well as recent experimental data from \citet{Bharadwaj2019}. Figure \ref{fig:StVRiHydraulicRadius}(a) shows that, once again, when we use a width- or diameter-based formulation for $\ell$, there is no single scaling relation that accurately describes all available data.  However, when the Strouhal and Richardson numbers are instead computed using the hydraulic radius, they all collapse along a single line. The least-squares fit to all data [not including the \citet{Bharadwaj2019} data, for reasons explained later] is $\mathrm{St}_{R_\mathrm{h}} = e^{-1.004}\mathrm{Ri}_{R_\mathrm{h}}^{0.406}$ with $r^2 = 0.984$, which is close to the scaling relation obtained for the simulation data alone [see Figure \ref{fig:StVRiTraditional}(b)].

Based on the current simulation and prior experimental results shown in Figure \ref{fig:StVRiHydraulicRadius}(b), as well as the similarity of the computed scaling coefficients to rational numbers, we propose the following shape-independent scaling relation between $\mathrm{St}_{R_\mathrm{h}}$ and $\mathrm{Ri}_{R_\mathrm{h}}$:
\begin{equation}\label{eq:law}
    \mathrm{St}_{R_\mathrm{h}} = e^{-1}\mathrm{Ri}_{R_\mathrm{h}}^{2/5}\,.
\end{equation}
The coefficient of correlation between this relation and the simulation and experimental data is $r^2 = 0.984$. The accuracy of the proposed relation compared to the data can also be determined by examining the residuals of the data with respect to the proposed fit, as shown in Figure~\ref{fig:StVRiHydraulicRadius}(c). For the proposed relation to be considered an acceptable fit to the data, the residuals should be scattered about zero and show no discernible trends with respect to the dependent variable (i.e., $\mathrm{Ri}_{R_\mathrm{h}}$). Figure~\ref{fig:StVRiHydraulicRadius}(c) shows that the residuals for prior experimental \citep{Hamins1992,Cetegen1996,Cetegen1998} and the present simulation data are all scattered about zero, indicating the lack of any persistent bias in Eq.\ \eqref{eq:law} with respect to this data. 

Rectangular inlet results from \citet{Bharadwaj2019} are also included in Figure~\ref{fig:StVRiHydraulicRadius}, but there is a consistent bias between this data and both the proposed scaling relation in Eq.\ \eqref{eq:law} and the least-squares fit in Figure~\ref{fig:StVRiHydraulicRadius}(b). This results in exclusively negative values of the residuals for data from the study by \citet{Bharadwaj2019}, as shown in Figure~\ref{fig:StVRiHydraulicRadius}(c). Thus, although the data from this study is quite precise, as indicated by the small scatter in the corresponding residuals in Figure~\ref{fig:StVRiHydraulicRadius}(c), the data appears to be inconsistent with other experimental and simulation results, as well as with the relation in Eq.\ \eqref{eq:law}. It is also worth noting that \citet{Bharadwaj2019} used an effective velocity instead of $V_0$ in their scaling relations, but in the present study this correction was found to be unnecessary for obtaining good agreement between the proposed relation in Eq.\ \eqref{eq:law} and all other simulation and experimental data.

\section{Conclusions\label{sec:conclusions}}
Adaptive mesh numerical simulations of buoyant jets have been performed for seven different inlet shapes: two circular inlets (corresponding to different diameters), three rectangular inlets (corresponding to different aspect ratios), an equilateral triangle inlet, and an annular inlet. A range of inlet velocities were simulated for each geometry, giving a range of inlet Richardson numbers for each case, and the resulting puffing Strouhal numbers were measured above the jet inlets.

Fundamental aspects of the flow evolution were found to be similar for each of the inlet shapes, including the buoyant rise of hot inlet gases, entrainment of cold ambient gases, pinching of the hot gases towards a central location above each inlet, formation of a KH instability along the interface between the hot and cold gases, and the subsequent pinch-off and shedding of vortical structures. This process was observed for each inlet shape, but the shed vortices were found to become less coherent as the complexity of the inlet increased. We found that, in general, the puffing frequency increased with inlet velocity and, for the rectangular cases, aspect ratio. 

Using width-based puffing Strouhal and inlet Richardson numbers, our results are consistent with prior experimental studies of round and planar jets, with the highest aspect ratio rectangular cases examined here closely corresponding to prior planar results. However, when using such width-based non-dimensional parameters, we did not find a single scaling relation that accurately describes the Strouhal-Richardson relationship for all inlet shapes. 

By contrast, if we instead use Strouhal and Richardson numbers based on the hydraulic radius, we do recover a single scaling relation, given in Eq.\ \eqref{eq:law}, that accurately describes the present computational results and prior planar and round experimental data. Recent experimental results for rectangular inlets \citep{Bharadwaj2019} were found to deviate from this scaling relation, but we also did not require any modification to the inlet velocity used to construct the Strouhal and Richardson numbers in order to obtain consistent agreement between the proposed scaling relation and all other data. The scaling relation in Eq.\ \eqref{eq:law} was found to closely describe both experimental and simulation results for Richardson numbers spanning four orders of magnitude. 

It is tempting to propose Eq.\ \eqref{eq:law} as a universal scaling relation for buoyant jet puffing that can be applied to any inlet shape. However, a broader range of inlet shapes, spanning a wider range of Richardson numbers, must be examined before this claim can be made with confidence. In particular, complex inlet shapes without any symmetries, or those with narrow connection points, require further study. The rational exponents in Eq.\ \eqref{eq:law} are also suggestive of a theoretically-based scaling relation, and further study may provide a rigorous justification for the proposed exponents. Finally, we found pronounced differences between results from \citet{Bharadwaj2019} and all other experimental and computational results. It is not clear why this discrepancy exists, particularly since the present rectangular results [simlar to the geometries studied by \citet{Bharadwaj2019}] are consistent with all other computational and experimental data. Consequently, this difference requires further study, and a resolution, in future work. 

\acknowledgments
Helpful discussions with Drs.\ Marcus S.~Day, Andrew Nonaka, and Werner J.A.~Dahm are gratefully acknowledged. NTW, GBR, and PEH were supported, in part, by the Strategic Environmental Research and Development Program under grant W912HQ-16-C-0026. CL was supported by the National Science Foundation Graduate Fellowship Program. Gift support from the 3M Company is also gratefully acknowledged.

\bibliographystyle{jfm}
\bibliography{references}

\end{document}